\documentclass[10pt,conference]{IEEEtran} 
\IEEEoverridecommandlockouts

\usepackage{cite}
\usepackage{amsmath,amssymb,amsfonts}
\usepackage{algorithmic}
\usepackage{graphicx}
\usepackage{textcomp}
\usepackage{xcolor}
\usepackage{xspace}
\usepackage{listings}

\lstdefinestyle{ath}{
  basicstyle=\ttfamily\scriptsize\color{bottlegreen},
  breaklines=true,          
  keepspaces=true,
  showstringspaces=false,
  columns=fullflexible,
  xleftmargin=0pt,
  aboveskip=2pt
}
\usepackage{xcolor}
\usepackage{tcolorbox}
\usepackage[hidelinks]{hyperref}

\usepackage{colortbl}
\usepackage{diagbox}

\newcommand{\sonar}{\textsc{Sonar}\xspace}

\def\BibTeX{{\rm B\kern-.05em{\sc i\kern-.025em b}\kern-.08em
    T\kern-.1667em\lower.7ex\hbox{E}\kern-.125emX}}
\usepackage{tikz}

\usetikzlibrary{positioning, fit, backgrounds, calc}
\usepackage{enumitem}
\usepackage{tabularx}
\usepackage{tcolorbox}
\usepackage{caption}
\usepackage{booktabs}
\usepackage{multirow}
\usepackage{makecell}
\usepackage{multicol}
\newtheorem{definition}{Definition}
\tcbset{
  mybox/.style={
    colframe=gray!70,
    colback=white,
    colbacktitle=green!10,  
    coltitle=black,        
    fonttitle=\bfseries\small,
    boxrule=0.5pt,
    arc=2pt,
    left=3pt,right=3pt,top=3pt,bottom=3pt,
    before skip=3pt, after skip=3pt,
    title filled          
  }
}
\tcbset{
  promptbox/.style={
    colframe=gray!70,
    colback=gray!10,
    boxrule=0.5pt,
    arc=2pt,
    left=3pt,right=3pt,top=3pt,bottom=3pt,
    before skip=3pt, after skip=3pt,
  }
}

\newtcolorbox{rqbox}{               
  colback=gray!10,          
  colframe=black,           
  boxrule=0.6pt,           
  arc=3mm,                 
  left=8pt,right=8pt,top=6pt,bottom=6pt, 
}

\newcommand{\ignore}[1]{}

\newcommand{\leak}[1]{{\setlength{\fboxrule}{0.4pt}\setlength{\fboxsep}{1pt}\fcolorbox{red}{blue!5}{\texttt{#1}}}}
\begin{document}

\title{SONAR: Task-Aware Code Summary Evaluation for LLM Consumers Without References}

\author{\IEEEauthorblockN{Simantika Bhattacharjee Dristi}
\IEEEauthorblockA{
\textit{University of Virginia}\\
Charlottesville, Virginia, USA \\
nwc8gr@virginia.edu}
\and
\IEEEauthorblockN{Matthew B. Dwyer}
\IEEEauthorblockA{
\textit{University of Virginia}\\
Charlottesville, Virginia, USA \\
matthewbdwyer@virginia.edu}

}

\maketitle

\begin{abstract}
Source code summaries have traditionally been evaluated from a human developer's perspective, with quality determined by how closely they resemble developer-written references and how well they align with human preferences. But this overlooks a growing reality: LLM-based tools and agents increasingly consume code summaries as inputs for software engineering (SE) tasks, and what makes a summary useful for a consuming agent on a task remains largely unexplored. 
\par To bridge this gap, we propose \sonar, a reference-free framework that evaluates source code summaries along four dimensions: \textit{Correctness}, \textit{Abstraction}, \textit{Conciseness}, and \textit{Fluency}. Rather than optimizing for a pre-written ``gold standard", \sonar introduces a novel code regeneration-based approach that uses a summary to regenerate code and leverages that reconstruction as a quality signal of the summary. This provides an empirical grounding that requires neither a reference summary nor the subjective judgment of humans or LLMs. 
\par We evaluate \sonar's dimensions on their ability to influence LLM performance across four downstream SE tasks. We find that \textit{Correctness}, followed by \textit{Abstraction}, significantly correlates with LLM performance, with correlations up to 14$\times$ higher than the best baseline. \textit{Conciseness} and \textit{Fluency}, though widely valued by human developers, remain mostly insignificant to an LLM consumer, suggesting that what makes a summary useful is task- and consumer-dependent. Through a large-scale evaluation of 11 popular LLMs using \sonar, we further identify the strengths and weaknesses of different models across each quality dimension, while offering insights to facilitate future research on task-aware summarization.
\end{abstract}


\section{Introduction}
Like many other software engineering (SE) tasks, source code summarization -- generating natural-language descriptions of code functionality and intent\cite{def} --
is increasingly driven by large language models (LLMs)~\cite{in_era_llm,calibration}.
Good summaries offer developers and automated agents a natural language anchor for reasoning about code by accurately and concisely describing behavior, while abstracting away implementation details. 
Although many techniques have been proposed for evaluating code summaries~\cite{bleu, simllm, side, meteor, rogue, in_era_llm, llm-as-ajudge, llm_judge3}, we argue that the true value of a summary should be judged in terms of the tasks it supports and the consumers it serves.

The growth of agentic software engineering~\cite{ahmed2025artificial} has reshaped
how software engineering tasks are executed and elevated agents as a principal
consumer of task-related information -- like summaries.
From this perspective, prior evaluation methods for
source code summaries share two significant limitations.
First, they \textbf{treat developer-written documentation as a gold standard.} From reference-based metrics such as BLEU~\cite{bleu}, ROUGE~\cite{rogue}, and METEOR~\cite{meteor} to the reference-free metric, SIDE~\cite{side}, existing evaluation techniques largely optimize toward developer-written documentation, despite prior evidence that it can be incomplete or ambiguous~\cite{badref1}. 
Second, they \textbf{assume human developers are the sole consumers of summaries} - an outdated assumption on which the entire evaluation paradigm rests. This was a reasonable assumption when summarization research first emerged, with the stated goal of helping developers understand unfamiliar code faster.  But the landscape has since shifted; LLM-based tools and agents now use code summaries as inputs for a wide range of downstream SE tasks~\cite{cross_language_retreival, test_oracle_gen1, code_tranlsation2, code_translation}. As a result, a developer-only view no longer reflects how summaries are used or how they should be evaluated.

These limitations are coupled: current techniques optimize summaries toward a flawed reference using metrics tailored to a class of consumers whose role in specific software engineering tasks is diminishing. 
To resolve these limitations, in this work,  we adopt a three-part strategy. First, we identify distinct quality dimensions across
which SE tasks may vary in their sensitivity.
Second, we introduce a reference-free evaluation framework that measures multiple summary dimensions using only the source code and the summary itself, without relying on any pre-existing documentation as a gold standard to match. 
Finally, rather than assessing how important each dimension is to human developers, we measure how they impact LLM performance in practice, shifting code summary evaluation toward an automated consumer's perspective.
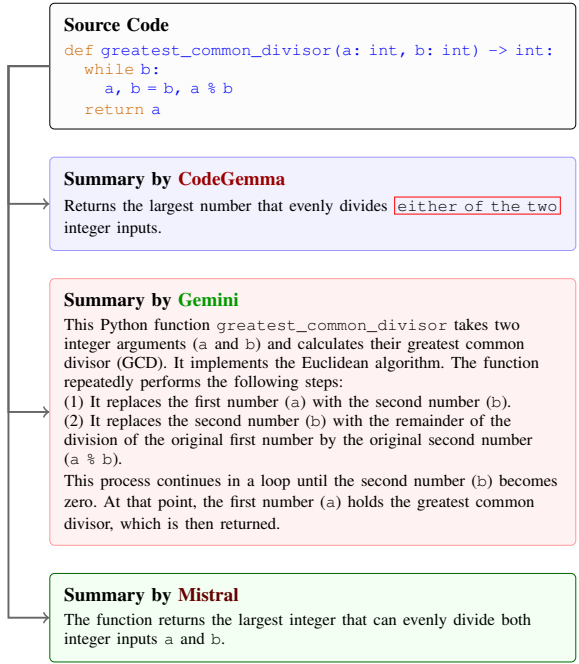
\begin{figure}[t]
\centering
\scalebox{0.9}{
\begin{tikzpicture}[
    font=\footnotesize,
    codebox/.style={draw, rounded corners=2pt, fill=gray!2,
                text width=7.3cm,
                align=left, inner sep=6pt},
    sumboxX/.style={draw=blue!40, rounded corners=2pt, fill=blue!5,
                text width=7.3cm,
                align=left, inner sep=6pt},
    sumboxY/.style={draw=red!40, rounded corners=2pt, fill=red!5,
                text width=7.3cm,
                align=left, inner sep=6pt},
    sumboxZ/.style={draw=green!40!black, rounded corners=2pt, fill=green!5,
                text width=7.3cm,
                align=left, inner sep=6pt},
    arrowM/.style={->,  thick, black!60}
]
\node[codebox, anchor=north] (code) at (0, 0) {
    \textbf{Source Code}\\[2pt]
    \scriptsize
    \texttt{\textcolor{orange!80!black}{def} \textcolor{blue}{greatest\_common\_divisor(a: int, b: int) -> int:}}\\
    \texttt{\phantom{xx}\textcolor{orange!80!black}{while} \textcolor{blue}{b:}}\\
    \texttt{\phantom{xxxx}\textcolor{blue}{a, b = b, a \% b}}\\
    \texttt{\phantom{xx}\textcolor{orange!80!black}{return} \textcolor{blue}{a}}
};
\node[sumboxX, anchor=north] (sumX) at ($(code.south) - (0, 0.4)$) {
    \textbf{Summary by \textcolor{red!60!black}{CodeGemma}}\\[2pt]
    \scriptsize
    Returns the largest number that evenly divides 
    \leak{either of the two} integer inputs.
};
\node[sumboxY, anchor=north] (sumY) at ($(sumX.south) - (0, 0.4)$) {
    \textbf{Summary by \textcolor{green!60!black}{Gemini}}\\[2pt]
    \scriptsize
    This Python function \texttt{greatest\_common\_divisor} takes two integer arguments 
    (\texttt{a} and \texttt{b}) and calculates their greatest common divisor (GCD).
    It implements the Euclidean algorithm. The function repeatedly performs the 
    following steps:\\
    (1) It replaces the first number (\texttt{a}) with the second number (\texttt{b}).\\
    (2) It replaces the second number (\texttt{b}) with the remainder of the 
    division of the original first number by the original second number (\texttt{a \% b}).\\
    This process continues in a loop until the second number (\texttt{b}) becomes zero. 
    At that point, the first number (\texttt{a}) holds the greatest common divisor, 
    which is then returned.
};
\node[sumboxZ, anchor=north] (sumZ) at ($(sumY.south) - (0, 0.4)$) {
    \textbf{Summary by \textcolor{red!40!black}{Mistral}}\\[2pt]
    \scriptsize
    The function returns the largest integer that can evenly divide both integer inputs \texttt{a} and \texttt{b}.
    
};
\coordinate (origin) at (code.west);
\draw[arrowM] (origin) -- ++(-0.6,0) |- (sumX.west);
\draw[arrowM] (origin) -- ++(-0.6,0) |- (sumY.west);
\draw[arrowM] (origin) -- ++(-0.6,0) |- (sumZ.west);
\end{tikzpicture}
}
\caption{LLM-generated summaries vary across quality dimensions under the same code and same prompt. Red borders denote incorrect information.}
\label{fig:motivating_example}
\vspace{-4mm}
\end{figure}
\par We begin with an observation that motivates our framework: for the same source code and prompt, different LLMs generate summaries that vary considerably in accuracy, reference to implementation detail, redundancy, and readability. For example, as Fig.~\ref{fig:motivating_example} shows, CodeGemma generates a summary that is concise and readable but contains inaccurate behavioral information. Gemini's summary, on the other hand, correctly describes the function, but includes many implementation details that tightly couple the summary to a single implementation of the underlying program behavior. Mistral offers a more balanced summary that accurately captures the program behavior, avoids unnecessary implementation detail, and remains concise and fluent enough to be readable.  These observations illustrate the multidimensional
nature of summary quality, leading us to identify four key dimensions of an LLM-generated code summary: \textbf{Correctness} - the extent to which it accurately captures the underlying code's behavior; \textbf{Abstraction} - the extent to which it avoids mentioning low-level implementation details; \textbf{Conciseness} - the extent to which it is free of redundant or verbose information; and \textbf{Fluency} - the extent to which it is expressed naturally and readably. 

To measure each of these dimensions in a principled way, we propose \sonar, a reference-free, multi-dimensional evaluation framework for source code summaries. \sonar is built on a novel code regeneration-based approach, where the summary is used to regenerate code(s), and the regenerated code(s) is then analyzed as a signal of summary quality. By going from code to summary and back to code, we leverage the causal relationship between a summary and the code(s) regenerated from it to empirically ground evaluation of code summaries, without relying on a reference, human, or LLM as a judge.

As described in \S\ref{sec:sonar}, we utilize the concept of round-trip correctness (RTC)~\cite{rtc} to measure three of the \sonar dimensions.
For \textit{Correctness}, given a summary and essential context, an independent code-generator LLM \textit{regenerates} code, and the degree of functional equivalence between the original and regenerated code determines the summary's correctness. 
For \textit{Abstraction}, we adapt RTC in a novel way to produce
a pool of regenerated implementations and then we measure the diversity of that pool. A more abstract summary allows code generators to explore a broader implementation space, resulting in a diverse set of regenerated implementations. 
For \textit{Conciseness}, we \textit{compress} the summary while retaining functional equivalence as determined by RTC and 
interpret the largest such degree of compression
as a measure of unnecessary information in the original summary. While accuracy (loosely analogous to correctness) and conciseness have previously been recognized as code summary quality aspects~\cite{llm-judge}, their evaluation has remained mostly text-centric with limited consideration of the functional behavior encoded in the summary. Abstraction, meanwhile, has been entirely overlooked in existing literature.

As we show through our experimental study in \S\ref{sec: Experimental Study}, \sonar addresses limitations of prior approaches.
First, \sonar effectively captures the quality dimensions of code summaries, achieving a 90\% overall agreement rate with ground-truth annotations across dimensions.
Second, at least one \sonar dimension significantly correlates with LLM performance across a set of downstream tasks, increasing the predictive signal by a factor of $1.3-14$ over the best baseline.  Moreover, the most predictive dimension varies by task, calling for a shift toward task-aware summarization. 
Finally, \sonar-based prompting provides a lightweight way to improve task-specific LLM summary generation, yielding up to a 10 percentage point (pp) gain in abstraction without any additional training.

The primary contributions of this paper lie in:
(1) introducing abstraction as a quality
attribute of code summaries that is relevant for some tasks;
(2) defining code regeneration-based measures for summary 
correctness, abstraction, and conciseness;
(3) defining \sonar, a reference-free, multi-dimensional evaluation framework using those measures; 
(4) demonstrating, through a large-scale evaluation
on downstream tasks, that different summary quality dimensions are important for different tasks; and
(5) showing that \sonar-guided prompting can improve LLM-generated summaries along task-relevant dimensions.

\section{Background and Related Work}
\subsection{Round-Trip-Correctness (RTC)}The core idea is to form a round trip from code to natural language and back to code. A natural language description is first generated from code, new code is synthesized from that description, and the resulting code is compared against the original. RTC has traditionally been used as an evaluation method for code LLMs~\cite{rtc, min2024accuracyevaluatingselfconsistencycode}.
\subsection{Evaluation of Code Summaries}
Reference-based methods have long dominated code summary evaluation~\cite{automated, in_era_llm}. These include exact match-based metrics, such as BLEU~\cite{bleu}, METEOR~\cite{meteor}, ROUGE~\cite{rogue}, that measure word overlap, and semantic match-based metrics, such as BERTScore~\cite{bertscore}, BLEURT~\cite{bleurt}, SentenceBERT~\cite{sentencebert}, SimLLM~\cite{simllm}, that assess semantic closeness to a reference. While widely used, these methods rely on developer-written documentation as reference summaries, which is not always available, and even when available, is often noisy or low quality~\cite{test_oracle_gen1, badref1, badref2}. Thus, the reference may be an unreliable oracle, causing high-quality summaries to be unfairly penalized when they differ from it.
\par Limited research has focused on developing reference-free evaluation methods for code summaries.  SIDE~\cite{side} reduces reference dependence at inference time, but its training continues to assume developer-written documentation as ground truth, thereby retaining a reference dependence. Moreover, SIDE measures quality only through embedding similarity between code and summary, rather than execution-based evidence. Therefore, its effectiveness in capturing true semantic alignment remains unclear.
\par A more recent line of work employs LLMs to judge code summary quality across dimensions such as content adequacy, coherence, fluency, understandability, and conciseness~\cite{llm-as-ajudge, llm-judge, llm_judge3}. However, these evaluations rely solely on the LLM’s internal reasoning about the summary text, with no empirical basis to justify the assigned scores. \sonar goes beyond all existing evaluation methods by eliminating the need for both training and reference summaries and by introducing regenerated code as a novel empirical anchor to ground the scores assigned to a summary under test.
\subsection{Code Summaries in Automated Workflows}
Natural language code summaries are playing an increasingly active role in automated software engineering workflows, from code translation and test generation to bug localization~\cite{code_tranlsation2, code_translation, test_gen, test_oracle_gen1, test_oracle_gen2, cross_language_retreival, bug_localization, bug_localiztaion2, webcodegen}. For example, Ahmad et al.~\cite{code_translation} leverage summaries as a pivot for unsupervised code translation, whereas Hossain et al.~\cite{test_oracle_gen1} use LLM-generated summaries as input to guide test oracle generation. Recent work also uses code summaries as proxy specifications for web code generation to evaluate the accessibility of LLM-generated code~\cite{webcodegen}. Despite these advances and evidence that code comments, and by extension, code summaries, can steer model behavior in meaningful ways~\cite{imani2025insideoutuncoveringcomment}, evaluation has remained entirely human-centric~\cite{simllm, side,in_era_llm, automated}, with no prior work assessing summaries from an LLM's perspective.

\section{The SONAR Approach}
\label{sec:sonar}
\begin{figure*}[t]
    \centering
  \includegraphics[width=0.95\textwidth]{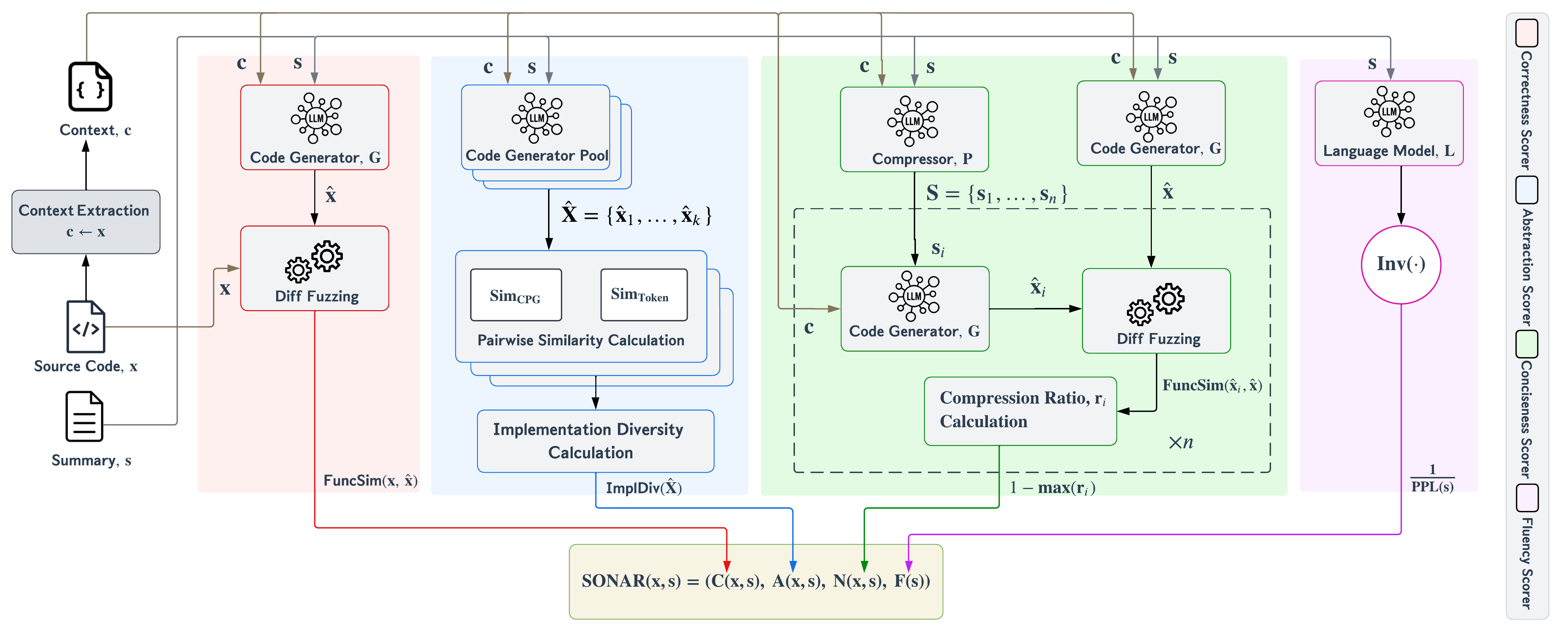}
   \captionsetup{margin={-0.60cm,0cm}}
    \caption{The SONAR Framework}
    \label{fig:SONAR}
\end{figure*}
\textsc{Sonar} is built on the principle of ``Regenerate, then Evaluate" - given a summary under test and its underlying source code, it first generates additional source code representation from the summary, and then probes those regenerated codes to infer the quality of the summary. Code regeneration provides a behavioral signal for summary quality. 
For example, if a code summary accurately captures the functional behavior of the underlying source code, then a code generator model, prompted with that summary, should be able to regenerate a functionally equivalent version of the source code. 
As another example, summary that is decoupled from low-level implementation details should allow a diverse pool of regenerated implementations, while one rich in such implementation details shrinks that space considerably. 

\subsection{Framework} 
\label{sec:frameork}
Figure \ref{fig:SONAR} presents an overview of the \sonar framework, which comprises four independent scorer modules, each measuring one quality dimension. Given a source code, $x$ and its corresponding summary $s$, \sonar returns four scores: 
\begin{equation*}
    \mathrm{SONAR}(x, s) =
    \left(C(x,s), A(x,s), N(x,s), F(s)\right)
\label{eq:SONAR}
\end{equation*}
where $C$ measures \textit{Correctness}, $N$ measures \textit{Conciseness}, $A$ measures \textit{Abstraction}, and $F$ measures \textit{Fluency}. Each score lies in $[0,1]$, where higher is better. In a pre-processing step,  the minimal code context, $c$, required to reconstruct the source code from the summary alone, is extracted from $x$. For function-level code, this context comprises the function signature - the function name, return type, and parameter names and types, along with any import statements the code depends on. The scorer modules share a code generator LLM, $G$, which takes the summary $s$ and extracted context $c$ as input and generates a candidate implementation:  

\begin{equation*}
    \hat{x} = G(s, c)
    \label{eq:generator}
\end{equation*}
This regenerated code $\hat{x}$ is analyzed by multiple \sonar modules to compute the quality scores. Each scorer module is discussed in detail in the sections that follow.
\subsubsection{Correctness Scorer}
\label{sec: Corr_Scorer}
This module quantifies correctness under the intuition that the more functionally similar $\hat{x}$ is to $x$, the more correct the summary.
\begin{definition}[Correctness]
Given source code $x$, summary $s$, and the regenerated code, $\hat{x}$, the correctness of $s$ with respect to $x$ is:
\begin{equation*}
    C(x, s) = \text{FuncSim}(x, \hat{x})
    \label{eq:correctness}
\end{equation*}
where $\text{FuncSim}(x, \hat{x})$ measures functional similarity.
\end{definition}
\par The scorer first instantiates $G$ with $s$ and $c$ to generate $\hat{x}$, and then computes functional similarity between $x$ and $\hat{x}$ using the differential fuzzing-based approach of Dristi et al.~\cite{local,Diff_Fuzz_Refac}. A test input set $T$ is generated on the fly, and both $x$ and $\hat{x}$ are run on each input. Functional similarity is defined as the ratio of inputs for which both produce identical outputs: 
\begin{equation}
\mathrm{FuncSim}(x,\hat{x}) =
\left|\{t \in T \mid x(t) = \hat{x}(t)\}\right|
/
|T|
\label{eq:FuncSim}
\end{equation}
Prior RTC work has either used unit test execution or reference-based metrics such as CodeBLEU~\cite{ren2020codebleumethodautomaticevaluation} as a functional similarity oracle. However, unit testing is infeasible in our setting since very few code-summary pairs in available datasets~\cite{CSN,FunCOM} come with associated test cases. Reference-based metrics, meanwhile, suffer from surface bias~\cite{local}, failing to reason about deeper code semantics. Our choice of differential fuzzing as the oracle resolves both issues: the test input set, $T$, is automatically generated using a byte-level fuzzer with no predefined tests needed, while execution-level comparison provides stronger evidence for the functional similarity score.  However, when the source code does not come with sufficient context to be run in isolation, reference-based code evaluation metrics can serve as a proxy for functional similarity.
\subsubsection{Abstraction Scorer} This module approximates abstraction in a code summary through the diversity of regenerated implementations: the more implementation details a summary encodes, the more constrained the regeneration space is, and the less diverse the regenerated implementations will be.
\begin{definition}[Abstraction]
Given source code $x$, summary $s$, and a set of regenerated implementations, $\hat{X}$, the abstraction of $s$ with respect to $x$ is:
\begin{equation*}
    A(x, s) = \text{ImplDiv}(\hat{X})
    \label{eq:abstraction_def}
\end{equation*}
where $\text{ImplDiv}(\hat{X})$ is the implementation-level diversity of $\hat{X}$.
\end{definition}
\par In this module, first, $\hat{X}$ is constructed using a pool of code generators $\mathcal{G} = \{G_1, \ldots, G_U\}$. Since different LLMs exhibit different structural biases and coding styles, cross-model sampling enables broader exploration of the implementation space than any single model could provide. $V$ implementations per code generator are further sampled at different temperatures to capture implementation diversity both within and across models. Given $s$, and $c$, this yields -
\begin{equation*}
    \hat{X} = \left\{\hat{x}_{i,j} \mid \hat{x}_{i,j} 
    \sim G_i(s, c)\right\}_{i=1,\ j=1}^{U,\ V}
    \label{eq:abstraction_pool}
\end{equation*}
a total of $U \times V$ regenerated implementations.
\par Whereas the correctness scorer (\S~\ref{sec: Corr_Scorer}) prompts $G$ without additional instruction, each code generator here is explicitly instructed to preserve any implementation-specific details stated in the summary, while making independent implementation choices only when the summary leaves them unspecified. Thus, abstract summaries allow the generators to explore a wider implementation space, while implementation-heavy ones constrain them and reduce diversity in $\hat{X}$. 
\par Next, the abstraction scorer computes the pairwise implementation-level similarity, $ImplSim$, between each pair of implementations in $\hat{X}$. Implementation details embedded in a summary can propagate to regenerated code at both structural and lexical levels. As shown in Fig. \ref{fig:abstraction_example}, summary (a) encodes specific data structures and variable names, which all regenerated implementations mimic. Summary (b), by contrast, describes the function behavior at a high level, allowing different regenerated implementations to use different variable names and data structures. Based on this observation, we decompose $ImplSim$ into two components: \textbf{Sim\textsubscript{CPG}}, which measures structural similarity between two codes using Code Property Graphs (CPGs) extracted via Joern~\cite{joern}, an open-source program analysis tool, and \textbf{Sim\textsubscript{Token}}, 
which measures lexical similarity through token-level overlap:
\begin{equation*}
    \text{ImplSim}(\hat{x}_i, \hat{x}_j) = 
    (\text{Sim}_{\text{CPG}}(\hat{x}_i, \hat{x}_j) + 
    \text{Sim}_{\text{Token}}(\hat{x}_i, \hat{x}_j))/{2}
    \label{eq:implsim}
\end{equation*}
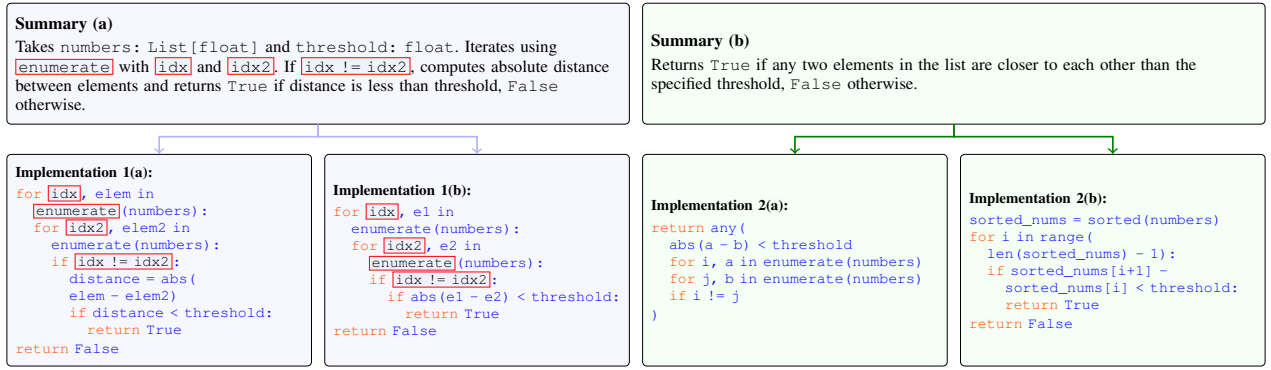
\begin{figure*}[t]
\centering
\scalebox{0.8}{
\begin{tikzpicture}[
    font=\footnotesize,
    sumbox/.style={draw, rounded corners=2pt,  
                text width=10cm, minimum height=2.0cm,
                align=left, inner sep=4pt},
    implbox/.style={draw, rounded corners=2pt, fill=green!4, 
                    text width=4.75cm, minimum height=3.5cm,
                    align=left, inner sep=4pt},
    leakbox/.style={draw, rounded corners=2pt,
                    fill=blue!3,
                    text width=4.75cm, minimum height=3.5cm,
                    align=left, inner sep=4pt},
    arrowA/.style={->, thick, blue!30},
    arrowB/.style={->, thick, green!50!black}
]

\node[sumbox, fill=blue!3, anchor=north west] (sumA) at (-10.5, 0) {
    \textbf{Summary (a)}\\[2pt]
    \footnotesize
    Takes \texttt{numbers: List[float]} and 
    \texttt{threshold: float}. Iterates using 
    \leak{enumerate} with \leak{idx} and \leak{idx2}. 
    If \leak{idx != idx2}, computes absolute distance 
    between elements and returns \texttt{True} if 
    distance is less than threshold, \texttt{False} 
    otherwise.
};

\node[sumbox, fill=green!4, anchor=north west] (sumB) at (0, 0) {
    \textbf{Summary (b)}\\[2pt]
    \footnotesize
    Returns \texttt{True} if any two elements in the list 
    are closer to each other than the specified 
    threshold, \texttt{False} otherwise.
};

\node[leakbox, anchor=north west] (implA1) at (-10.5, -2.5) {
    \scriptsize\textbf{Implementation 1(a):}\\[2pt]
    \texttt{\textcolor{orange!60!red}{for} \textcolor{blue}{}}\leak{idx}\texttt{\textcolor{blue}{, elem in}}\\
    \texttt{\phantom{xx}}\leak{enumerate}\texttt{\textcolor{blue}{(numbers):}}\\
    \texttt{\phantom{xx}\textcolor{orange!60!red}{for} \textcolor{blue}{}}\leak{idx2}\texttt{\textcolor{blue}{, elem2 in}}\\
    \texttt{\phantom{xxxx}\textcolor{blue}{enumerate(numbers):}}\\
    \texttt{\phantom{xxxx}\textcolor{orange!60!red}{if} \textcolor{blue}{}}\leak{idx != idx2}\texttt{\textcolor{blue}{:}}\\
    \texttt{\phantom{xxxxxx}\textcolor{blue}{distance = abs(}}\\
    \texttt{\phantom{xxxxxx}\textcolor{blue}{elem - elem2)}}\\
    \texttt{\phantom{xxxxxx}\textcolor{orange!60!red}{if} \textcolor{blue}{distance < threshold:}}\\
    \texttt{\phantom{xxxxxxxx}\textcolor{orange!60!red}{return} \textcolor{blue}{True}}\\
    \texttt{\textcolor{orange!60!red}{return} \textcolor{blue}{False}}
};
\node[leakbox, anchor=north west] (implA2) at (-5.25, -2.5) {
    \scriptsize\textbf{Implementation 1(b):}\\[2pt]
    \texttt{\textcolor{orange!60!red}{for} \textcolor{blue}{}}\leak{idx}\texttt{\textcolor{blue}{, e1 in}}\\
    \texttt{\phantom{xx}\textcolor{blue}{enumerate(numbers):}}\\
    \texttt{\phantom{xx}\textcolor{orange!60!red}{for} \textcolor{blue}{}}\leak{idx2}\texttt{\textcolor{blue}{, e2 in}}\\
    \texttt{\phantom{xxxx}}\leak{enumerate}\texttt{\textcolor{blue}{(numbers):}}\\
    \texttt{\phantom{xxxx}\textcolor{orange!60!red}{if} \textcolor{blue}{}}\leak{idx != idx2}\texttt{\textcolor{blue}{:}}\\
    \texttt{\phantom{xxxxxx}\textcolor{orange!60!red}{if} \textcolor{blue}{abs(e1 - e2) < threshold:}}\\
    \texttt{\phantom{xxxxxxxx}\textcolor{orange!60!red}{return} \textcolor{blue}{True}}\\
    \texttt{\textcolor{orange!60!red}{return} \textcolor{blue}{False}}\\[2pt]
};
\node[implbox, anchor=north west] (implB1) at (0, -2.5) {
    \scriptsize\textbf{Implementation 2(a):}\\[2pt]
    \texttt{\textcolor{orange!60!red}{return} \textcolor{blue}{any(}}\\
    \texttt{\phantom{xx}\textcolor{blue}{abs(a - b) < threshold}}\\
    \texttt{\phantom{xx}\textcolor{orange!60!red}{for} \textcolor{blue}{i, a in enumerate(numbers)}}\\
    \texttt{\phantom{xx}\textcolor{orange!60!red}{for} \textcolor{blue}{j, b in enumerate(numbers)}}\\
    \texttt{\phantom{xx}\textcolor{orange!60!red}{if} \textcolor{blue}{i != j}}\\
    \texttt{\textcolor{blue}{)}}
};
\node[implbox, anchor=north west] (implB2) at (5.25, -2.5) {
    \scriptsize\textbf{Implementation 2(b):}\\[2pt]
    \texttt{\textcolor{blue}{sorted\_nums = sorted(numbers)}}\\
    \texttt{\textcolor{orange!60!red}{for} \textcolor{blue}{i in range(}}\\
    \texttt{\phantom{xx}\textcolor{blue}{len(sorted\_nums) - 1):}}\\
    \texttt{\phantom{xx}\textcolor{orange!60!red}{if} \textcolor{blue}{sorted\_nums[i+1] -}}\\
    \texttt{\phantom{xxxx}\textcolor{blue}{sorted\_nums[i] < threshold:}}\\
    \texttt{\phantom{xxxx}\textcolor{orange!60!red}{return} \textcolor{blue}{True}}\\
    \texttt{\textcolor{orange!60!red}{return} \textcolor{blue}{False}}
};

\draw[arrowA] (sumA.south) -- ++(0,-0.2) -| (implA1.north);
\draw[arrowA] (sumA.south) -- ++(0,-0.2) -| (implA2.north);
\draw[arrowB] (sumB.south) -- ++(0,-0.2) -| (implB1.north);
\draw[arrowB] (sumB.south) -- ++(0,-0.2) -| (implB2.north);

\end{tikzpicture}
}
\caption{Implementation details in a summary affect the diversity of regenerated implementations. Red borders highlight implementation-specific details propagated from the summary to the regenerated codes.}
\label{fig:abstraction_example}
\end{figure*}
The implementation-level diversity of $\hat{X}$ is then calculated as the complement of the average pairwise similarity across all implementations in $\hat{X}$:
\begin{equation*}
    \text{ImplDiv}(\hat{X}) = 1 - \frac{2}{k(k-1)} 
    \sum_{i < j} \text{ImplSim}(\hat{x}_i, \hat{x}_j)
    \label{eq:imp_div}
\end{equation*}
where $k = U \times V$. While prior work has largely relied on Abstract Syntax Trees (ASTs) for structural comparison~\cite{AST_1, AST_2, local}, CPGs subsume ASTs by additionally capturing control flow and data dependencies, providing a more discriminative structural similarity measure. For each implementation, we extract its CPG and represent it as a set of labeled edges. Each edge in this set connects two program elements (statements, variables, etc.) and is labeled by its relation type (e.g., AST, control-flow, or data-dependency). We compute Sim\textsubscript{CPG} as the Jaccard similarity between the two CPG edge sets. Since Jaccard similarity is commonly used in code analysis and clone detection~\cite{jaccard_1, jaccard_2}, we also use it for Sim\textsubscript{Token}, though the proposed approach to measure abstraction can be used with other similarity measures as well. To ensure $\text{Sim}_{\text{token}}$ captures meaningful implementation overlap, we exclude comments and docstrings, as they do not represent implementation content. We also discard language-specific keywords and tokens appearing in the context, $c$, since their overlap across implementations is unavoidable and therefore uninformative.

\subsubsection{Conciseness Scorer}While conciseness is traditionally assessed by measuring redundancy, determining what is redundant in a code summary is non-trivial. This module takes a behavioral approach - content is redundant if its removal still allows functionally equivalent code to be regenerated. Redundancy is thus reflected in how much a summary can be compressed while preserving behavior.

\begin{definition}[Conciseness]
Given source code $x$, summary $s$, and a set of compressed summaries $\mathcal{S} = \{s_1, \ldots, s_n\}$ derived from $s$, let $\hat{x} = G(s, c)$ be the code regenerated from the original summary and $\hat{x}_i = G(s_i, c)$ be the code regenerated from each $s_i \in \mathcal{S}$. The conciseness of $s$ with respect to $x$ is:
\begin{equation}
    N(x, s) = 1 - \max_{i \in \{1,\ldots,n\}} r_i
    \label{eq:conciseness}
\end{equation}
where $r_i$ is the compression ratio of $s_i$, defined as:
\begin{equation}
    r_i = \begin{cases} 
    1 - \frac{|s_i|}{|s|} & \text{if } 
    \text{FuncSim}(\hat{x}_i, \hat{x}) = 1.0 \\ 
    0 & \text{otherwise} 
    \end{cases}
    \label{eq:compression_ratio}
\end{equation}
\end{definition}

\par The conciseness scorer starts by generating $\hat{x}$ from $s$  using $G$. At the same time, a compressor LLM, $P$, is prompted $n$ times independently to compress $s$ while preserving all behavioral information, producing the set of compressed summaries, $\mathcal{S}$. Each $s_i \in \mathcal{S}$ is used to generate  $\hat{x}_i$ via $G$, which is evaluated for functional similarity with $\hat{x}$ using Eq.~\ref{eq:FuncSim}. Next, the compression ratio $r_i$ (Eq.~\ref{eq:compression_ratio}) is calculated to assess how much of $s$ can be removed while preserving behavior. A larger $r_i$ reflects greater redundancy, while $r_i = 0$ indicates that $s_i$ failed to preserve 
behavior. The conciseness score is then denoted by the complement of the maximum compression ratio in $\mathcal{S}$ (Eq.~\ref{eq:conciseness}). A summary that cannot be compressed further without losing behavior has all $r_i = 0$ and therefore receives a score of $1.0$. On the other hand, the more redundancy a summary carries, the greater behavior-preserving compression it allows, and the lower its conciseness score. We use $n$ compressions rather than one to get a more reliable conciseness estimate. Multiple compressions increase the likelihood of finding the tightest behavior-preserving compression. Conversely, if none of the $n$ compressed summaries preserve behavior, this provides stronger evidence that the original summary is already concise.

\subsubsection{Fluency Scorer} Fluency is the only \sonar dimension that treats the summary as pure natural language, independent of the source code. With prior works consistently highlighting fluency as a property valued by developers~\cite{in_era_llm, calibration, automated, side}, its inclusion in \sonar allows us to investigate whether LLM agents and human developers prioritize the same aspects of summary quality. Since code regeneration yields no useful signal for a purely textual property, this module follows the established practice of measuring fluency as the inverse of language model perplexity~\cite{gpt2_perplexity, gpt2_perp_2}.
\begin{definition}[Fluency]
Given summary $s$ and language model $L$, the fluency of $s$ is defined as: $F(s) = 1/\text{PPL}(s)$
where $\text{PPL}(s)$ denotes the perplexity of $s$ under $L$.
\end{definition}

\begin{figure}[!t]
\centering
\scalebox{0.9}{
\begin{tcolorbox}[mybox]
\begin{tcolorbox}[mybox, title={Vanilla Prompt}]
\footnotesize
\textbf{Task:} Summarize the following Python function in 
natural language.\\
\textbf{Instructions:}\\
-~Output ONLY the summary.\\
-~Do NOT include any explanations, comments, or markdown\\
\textbf{Python function:} $\langle$\textit{Source Code}$\rangle$
\end{tcolorbox}
\vspace{0.6em}
\begin{tcolorbox}[mybox, title={Dimension-Aware Prompt}]
\footnotesize
\textbf{Task:} Summarize the following Python function in 
natural language, such that your summary satisfies these 
four qualities:\\
-~\textbf{Correctness:} accurately describe the behavior 
of the function.\\
-~\textbf{Abstraction:} avoid low-level implementation details.\\
-~\textbf{Conciseness:} avoid redundancy.\\
-~\textbf{Fluency:} be grammatically correct, and naturally readable.\\
\textbf{Instructions:}\\
-~Output ONLY the summary.\\
-~Do NOT include any explanations, comments, or markdown\\
\textbf{Python function:} $\langle$\textit{Source Code}$\rangle$
\end{tcolorbox}
\vspace{0.6em}
\begin{tcolorbox}[mybox, title={Few-Shot Prompt}]
\footnotesize
\textbf{Task:} Summarize the following Python function in 
natural language.\\
\textbf{Instructions:}\\
-~Output ONLY the summary.\\
-~Do NOT include any explanations, comments, or markdown\\
The following examples illustrate the expected output:\\
\textbf{Example 1:}\\
Python function: $\langle$\textit{Example Source Code 1}$\rangle$\\
Summary: $\langle$\textit{Example Summary 1}$\rangle$\\
\begin{tikzpicture}
\draw[dashed] (0,0) -- (3,0);
\end{tikzpicture}
\textbf{\\Example N:}\\
Python function: $\langle$\textit{Example Source Code N}$\rangle$\\
Summary: $\langle$\textit{Example Summary N}$\rangle$\\
\textbf{Now summarize this function:}\\
\textbf{Python function:} $\langle$\textit{Source Code}$\rangle$
\end{tcolorbox}
\end{tcolorbox}
}
\caption{Prompt templates used for summary generation.}
\label{fig:prompts}
\end{figure}

\section{Experimental Study}
\label{sec: Experimental Study}
Our experimental study is structured around \textit{four} research questions : \\
\noindent\textbf{RQ1:} How reliably does \sonar capture different quality dimensions of a code summary?\\
\noindent\textbf{RQ2:} How do \sonar's quality dimensions correlate with LLM performance in downstream software engineering tasks?\\
\noindent\textbf{RQ3:}  How do different LLMs compare in terms of summary quality when evaluated using \sonar? \\
\noindent\textbf{RQ4:} How sensitive is \sonar to the choice of LLM across the framework's modules?

\par To address these questions, we generate summaries using LLMs, score them with \sonar, and then analyze the summaries and their corresponding scores. To note, all summaries analyzed in this paper are LLM-generated; while manual curation could yield greater within-dimension variance, it would not reflect how summarization works in practice. This section outlines the summary generation setup, the implementation details of \sonar, and the evaluation results and analysis.
\subsection{Summary Generation} 
\label{sec:Summary_GEn}
\subsubsection{Datasets} We select the datasets based on two criteria. First, each dataset must include source code paired with human-written comments, docstrings, or prompts that can serve as reference summaries for comparison with baselines (\S~\ref{sec:rq2}). Second, the code must be executable, either standalone or with sufficient context, to support differential fuzzing for \textit{Correctness\footnote{\textit{Italicized} dimension names (e.g., \textit{Correctness}) denote \sonar's measurement of that dimension; non-italicized usage denotes the general concept.}} and \textit{Conciseness} scoring (Eq.\ref{eq:FuncSim}), and downstream task evaluation (\S~\ref{sec:rq2}). However, almost every widely used code summarization dataset~\cite{FunCOM, TL_Code_Sum, edinburg_NLP, CSN}, while providing reference documentation, contains non-executable code. This gap can be attributed to the text-only nature of existing summary evaluation, where no prior work has used execution-based validation to evaluate summaries. We therefore use popular coding benchmarks instead: HumanEval~\cite{HumanEval}, MBPP~\cite{MBPP}, BigCodeBench~\cite{BCB}, and The Vault~\cite{the_Vault}, all of which meet both selection criteria. We construct a pool of 500 functions, comprising all 164 from HumanEval and 112 sampled from each remaining dataset, for summary generation and subsequent evaluation. While HumanEval and MBPP include simple standalone functions, BigCodeBench and The Vault contain more complex programming tasks, together providing a function pool with varying levels of code complexity. \textit{Code} and \textit{source code} are used interchangeably throughout this paper to refer to function-level Python code.

\subsubsection{Large Language Models (LLMs)} As code summarizers, we select 11 popular LLMs spanning a range of model families and sizes (Fig.~\ref{fig:RQ3_Spider_chart}). These include large closed-source models such as Gemini 2.5 Flash and Kimi K2.5, medium-scale open-source models like CodeLlama-13B and Mistral-7B, and smaller models like Phi-3.5 Mini and  CodeT5 (220M), a compact Seq2Seq model fine-tuned specifically on the summarization task. For summary generation, all models are run with identical decoding parameters and temperature settings.

\subsubsection{Prompting Strategies} 
\label{Section:prompts}
Of our selected models, 9 are instruction-tuned, excluding CodeT5 and StarCoder2. For the instruction-tuned models, we use three prompting strategies to generate summaries (Fig.~\ref{fig:prompts}): Vanilla Prompting, which simply asks the model to summarize a function with no additional guidance; Dimension-Aware Prompting, which explicitly instructs the model to satisfy \sonar's four quality dimensions; and Few-Shot Prompting, which augments the prompt with 2 examples to steer the model toward desired summaries.
\subsection{Implementation Details of \sonar} To implement \sonar (See \S~\ref{sec:frameork} for notation), we use Gemini-2.5 Flash as $G$ in the Correctness Scorer. In the Abstraction Scorer, we use $U=3$ generators in $\mathcal{G}$: Gemini-2.5 Flash, Qwen-2.5-Coder-32B, and OLMo-32B-Instruct, all known for their strong code-generation ability~\cite{olmo, lmmarketcap_bigcodebench}. From each model, we sample $V=3$ implementations, yielding a pool of 9 regenerated implementations for \textit{Abstraction} calculation. Gemini-2.5 Flash serves as both the compressor, $P$, and generator, $G$, in the Conciseness Scorer, with $n=3$ summaries forming $\mathcal{S}$. Finally, following prior work~\cite{gpt2_perp_2}, we use GPT-2 as the language model, $L$, in the Fluency Scorer.
\subsection{RQ1: Reliability Assessment of \sonar's Scores}
\label{sec:rq1}
\begin{table}[t]
\centering
\scriptsize
\setlength{\tabcolsep}{4pt}
\renewcommand{\arraystretch}{1.3}
\caption{Agreement rate (\%): SONAR vs LLM-only Scorers.}
\label{tab:rq1_table}
\begin{tabular}{lcccc}
\toprule
\textbf{Scorer} & \textbf{Correctness} & \textbf{Abstraction} & \textbf{Conciseness} & \textbf{Overall} \\
\midrule
SONAR           & \textbf{96} & \textbf{87} & \textbf{87} & \textbf{90} \\
Gemini 3.1 Pro  & 51 & 76 & 62 & 63 \\
Claude Opus 4.8 & 60 & 77 & 79 & 72\\
\bottomrule
\end{tabular}
\end{table}
\subsubsection{Experimental Setup}  
This experiment evaluates the extent to which each \sonar dimension captures the quality aspect it is designed to measure. Since \textit{Fluency} follows prior work, we focus on the dimensions for which \sonar proposes new measurement methods: Correctness, Abstraction, and Conciseness. Because existing works do not empirically evaluate these dimensions~\cite{docstringeval}, establishing a reliable ground truth for validation remains challenging. We therefore validate these dimensions using human preference labels. For each dimension, we sample 100 pairs of summaries $(A,B)$, where both summaries belong to the same code. Two authors independently annotate each pair for the target dimension by choosing “A,” “B,” or “Not Sure,” indicating which summary is more correct, more abstract, or more concise. Only pairs where both annotators unanimously agree on a preference are retained as ground truth labels. This results in 84 Correctness pairs, 86 Abstraction pairs, and 89 Conciseness pairs. We compute agreement rate as the percentage of pairs where \sonar assigns higher score to the human-preferred summary. For comparison, we use two independent LLM-based scorers as baselines: Gemini 3.1 Pro and Claude Opus 4.8, both ranking within the top 4\% of reasoning models as of mid-2026~\cite{llmstats_reasoning_2026}. Each LLM scorer receives a code, its summary, and a clear definition of the target dimension, and is asked to assign a score between 0 and 1 for that dimension. Agreement rates of \sonar and the baselines are shown in Table~\ref{tab:rq1_table}.
\subsubsection{Results and Analysis} 
\sonar achieves \textbf{87--96\%} agreement across dimensions and \textbf{90\%} overall (Table~\ref{tab:rq1_table}), compared to the best baseline, at 72\%. This \textbf{18 pp} improvement over the best baseline underscores \sonar's reliability in capturing the intended quality dimension. The highest agreement is for correctness at 96\%, where \sonar benefits from execution-based validation. In contrast, the LLM-only scorers are confined to source code and summary text alone, and hence struggle most with correctness. Although abstraction and conciseness are more perceptible from the summary text, \sonar still outperforms the best baseline by 10 and 8 pp, respectively. Notably, \sonar is implemented with Gemini-2.5 Flash, which ranks well below the top-tier baseline models. This suggests that \sonar’s advantage comes from its design rather than a stronger underlying model; we further test this in \S~\ref{sec:rq4}.

\begin{rqbox}
\textbf{Findings of RQ1: }\sonar reliably captures the intended quality dimensions of code summaries, achieving a 90\% overall agreement rate with ground-truth annotations.
\end{rqbox}

\subsection{RQ2: Correlation Analysis between \sonar's Dimensions and Downstream SE Task Performance}
\label{sec:rq2}
\begin{table}[t]
\centering
\scriptsize
\setlength{\tabcolsep}{4.5pt}
\renewcommand{\arraystretch}{1.2}
\caption{Experimental setup for each downstream SE task. ``n'' denotes the number of summaries used per task. ``Qwen", ``OLMo", ``Gemini" refer to Qwen2.5-Coder-32B, OLMo-32B-Instruct, and Gemini-2.5-Flash, respectively.}
\label{tab:task_setup}
\begin{tabular}{>{\raggedright\arraybackslash}p{2.4cm}>{\raggedright\arraybackslash}p{2.0cm}>{\raggedright\arraybackslash}p{2.5cm}>{\raggedright\arraybackslash}l}
\toprule
\textbf{Task} & \textbf{Metric} & \textbf{Models} & \textbf{n} \\
\midrule
Retrieval & MRR~\cite{CSN} & UniXcoder, CodeT5+, GraphCodeBERT & 1414 \\
\arrayrulecolor{gray!80}\hline\arrayrulecolor{black}
\multirow{2}{*}{Translation} & CA@1~\cite{trans_diversity}, & \multirow{2}{*}{Qwen, OLMo, Gemini} & \multirow{2}{*}{1414} \\
 & Edit Distance~\cite{edit_distance} & & \\
\arrayrulecolor{gray!80}\hline\arrayrulecolor{black}
\multirow{2}{*}{Optimization} & Pass@1~\cite{HumanEval}, & \multirow{2}{*}{Qwen, OLMo, Gemini} & \multirow{2}{*}{1085} \\
 & Speedup~\cite{speedup} & & \\
\arrayrulecolor{gray!80}\hline\arrayrulecolor{black}
Test Oracle Generation & Success Rate~\cite{TOGLL} & Qwen, OLMo, Gemini & 2152 \\
\bottomrule
\end{tabular}
\end{table}

\begin{table*}[t]
\centering
\scriptsize
\setlength{\tabcolsep}{4pt}
\renewcommand{\arraystretch}{1.35}
\caption{$r_{\rho}$ between each method and downstream task performance. $^{*}p<0.05$, $^{**}p<0.01$, $^{***}p<0.001$; unmarked is non-significant ($p \geq 0.05$). Bold values indicate the highest significant correlation per task metric.}
\label{tab:rq2_table}
\begin{tabular}{ll|l|l|l|l|l|l}
\toprule
\multicolumn{2}{c|}{\multirow{2}{2.0cm}{\diagbox[width=2.2cm, height=1.0cm]{\textbf{Method}}{\textbf{Task}}}} 
& \textbf{Retrieval} & \multicolumn{2}{c|}{\textbf{Translation}} & \multicolumn{2}{c|}{\textbf{Optimization}} & \textbf{Test Oracle Gen.} \\
\cmidrule(lr){3-3} \cmidrule(lr){4-5} \cmidrule(lr){6-7} \cmidrule(lr){8-8}
& & MRR & CA@1 & Edit Distance & Pass@1 & Speedup & Success Rate \\
\midrule
\multirow{4}{*}{\rotatebox{90}{\textbf{SONAR}}}
& Correctness & $+0.06(\pm 0.04)$ & $\mathbf{+0.41(\pm 0.07)}^{\boldsymbol{***}}$ & $-0.14(\pm 0.02)^{\boldsymbol{***}}$ & $\mathbf{+0.59(\pm 0.05)}^{\boldsymbol{***}}$ & $-0.05(\pm 0.04)$ & $\mathbf{+0.41(\pm 0.06)}^{\boldsymbol{***}}$ \\
& Abstraction & $\mathbf{+0.18(\pm 0.04)}^{\boldsymbol{***}}$ & $-0.11(\pm 0.05)^{\boldsymbol{*}}$ & $\mathbf{+0.28(\pm 0.05)}^{\boldsymbol{***}}$ & $-0.02(\pm 0.01)$ & $\mathbf{+0.11(\pm 0.01)}^{\boldsymbol{**}}$ & $-0.21(\pm 0.04)^{\boldsymbol{***}}$ \\
& Conciseness & $+0.06(\pm 0.08)$ & $+0.04(\pm 0.03)$ & $+0.05(\pm 0.02)$ & $-0.01(\pm 0.02)$ & $-0.01(\pm 0.03)$ & $+0.03(\pm 0.02)$ \\
& Fluency     & $+0.03(\pm 0.06)$ & $-0.02(\pm 0.04)$ & $-0.07(\pm 0.04)^{\boldsymbol{*}}$ & $+0.06(\pm 0.03)$ & $-0.06(\pm 0.03)$ & $+0.05(\pm 0.03)$ \\
\arrayrulecolor{gray!80}\hline\arrayrulecolor{black}
\multirow{5}{*}{\rotatebox{90}{\textbf{Baselines}}}
& BLEU-4      & $+0.02(\pm 0.08)$ & $-0.09(\pm 0.02)^{\boldsymbol{*}}$ & $+0.07(\pm 0.02)$ & $+0.07(\pm 0.02)$ & $+0.06(\pm 0.04)$ & $+0.09(\pm 0.01)^{\boldsymbol{**}}$ \\
& METEOR      & $+0.02(\pm 0.06)$ & $-0.01(\pm 0.03)$ & $+0.03(\pm 0.01)$ & $+0.14(\pm 0.01)^{\boldsymbol{***}}$ & $-0.05(\pm 0.06)$ & $+0.12(\pm 0.03)^{\boldsymbol{***}}$ \\
& ROUGE-L     & $+0.01(\pm 0.08)$ & $-0.11(\pm 0.04)^{\boldsymbol{*}}$ & $+0.01(\pm 0.03)$ & $-0.06(\pm 0.01)$ & $+0.09(\pm 0.06)$ & $+0.09(\pm 0.03)^{\boldsymbol{**}}$ \\
& BLEURT      & $+0.14(\pm 0.07)^{\boldsymbol{*}}$ & $+0.02(\pm 0.07)^{\boldsymbol{*}}$ & $-0.02(\pm 0.05)$ & $+0.15(\pm 0.02)^{\boldsymbol{***}}$ & $-0.03(\pm 0.05)$ & $+0.20(\pm 0.03)^{\boldsymbol{***}}$ \\
& BERTScore   & $+0.08(\pm 0.08)$ & $-0.04(\pm 0.03)$ & $+0.00(\pm 0.04)$ & $+0.08(\pm 0.03)$ & $-0.00(\pm 0.04)$ & $+0.12(\pm 0.02)^{\boldsymbol{***}}$ \\
\bottomrule
\end{tabular}
\end{table*}

\subsubsection{Task Selection} To evaluate the impact of \sonar's dimensions on LLM performance, we focus on prior work that has used code summaries, comments, or documentation as input for some SE task, and select four representative tasks that collectively capture distinct downstream uses of code summaries: code retrieval~\cite{cross_language_retreival, cross_language_retreival_2, cross_lang_retreival3}, code translation~\cite{code_tranlsation2, code_translation}, code optimization~\cite{codeopt1, codeopt2}, and test oracle generation~\cite{test_oracle_gen1, test_oracle_gen2}. Since the selected tasks require test cases or cross-language equivalents, we restrict our experiments to HumanEval and MBPP, which provide these artifacts. We randomly sample 50 source codes from each dataset and collect all summaries generated for them from the pool described in \S~\ref{sec:Summary_GEn}. For each task, we retain only summaries whose source code includes the artifacts required for that task, leading to varying sample sizes across tasks. Table \ref{tab:task_setup} reports the sample size, evaluation metrics, and models used per task. Zero-shot prompts are used for all tasks to avoid prompt-specific effects. \\
\textbf{Code Retrieval.} Given a summary as the query, the task is to retrieve the most relevant code by matching the summary's embedding against the embeddings of codes in a candidate pool. Since cross-language retrieval is particularly challenging~\cite{cross_language_retreival}, we focus on the Python-to-Java setting, where a Python function summary is used to retrieve its Java equivalent. We use the same retrieval pool as~\cite{cross_language_retreival}, augmented with the sampled HumanEval and MBPP  codes and their equivalent Java implementations. We report MRR~\cite{CSN} as the metric, which in our per-summary setting reduces to RR. \\
\textbf{Code Translation.} Given a Python function summary as specification, an LLM is prompted to implement the same functionality in Java. Alongside translation accuracy, we measure the average edit distance (the diversity metric of~\cite{trans_diversity}) among 5 independently sampled Java implementations from the same LLM, since diverse translations are known to benefit data augmentation~\cite{trans_diversity} and runtime efficiency~\cite{trans_diversity2}.\\
\textbf{Code Optimization.} The LLM is tasked with generating the most runtime-efficient implementation of the functionality described by the summary. Execution speedup is measured using the large-input stress tests and experimental setup of~\cite{speedup}.\\
\textbf{Test Oracle Generation} Building on~\cite{test_oracle_gen1}, we provide the LLM with the function signature of the method under test (MUT), along with the corresponding summary. We then ask the LLM to generate 10 assert statements to test the MUT. 
\subsubsection{Baselines} Baselines include five popular reference-based metrics: text-based, BLEU-4, METEOR, ROUGE-L, and embedding-based, BLEURT, and BERTScore (F1). Although SIDE~\cite{side} is a relevant reference-free baseline, its Java-specific training makes comparison with \textsc{Sonar}’s Python implementation infeasible here. For the baselines, we use HumanEval and MBPP's natural-language descriptions as references, since they are human-written~\cite{prompt_issue} and thus provide reasonable gold standards for comparison.

\subsubsection{Evaluation Metric}We use \textbf{Spearman’s rank correlation coefficient}~\cite{spearman1904}, $\boldsymbol{r}_{\rho}$, to measure the correlation between \textsc{Sonar}’s dimensions and downstream task performance. Spearman is appropriate because the dimension scores and task metrics are not normally distributed (validated with the Shapiro-Wilk normality test \cite{shapiro_test}), span different value ranges, and are not assumed to be linearly related. Instead, we are interested in whether changes in scores are monotonically associated with increases or decreases in downstream performance, which Spearman captures through rank-based correlation. To account for model-specific variation, we evaluate three LLMs per task, compute $r{\rho}$ separately for each LLM and metric, and aggregate correlations using Fisher-$z$ averaging~\cite{fisherZ}. Table~\ref{tab:rq2_table} reports the $\text{mean}(\pm\text{standard deviation})$ of $r_{\rho}$, with the corresponding significance level.

\subsubsection{Results and Analysis}
As shown in Table~\ref{tab:rq2_table}, \textbf{at least one \sonar dimension significantly and positively correlates with every task metric, with up to a \textbf{14 times} improvement in predictive strength over the strongest baseline.} For code retrieval, only \textit{Abstraction} shows significant correlation $(r_{\rho}= 0.18)$, suggesting that summaries with fewer implementation details generalize better across languages and more effectively support cross-language retrieval. Since retrieval relies solely on embedding similarity and does not verify functional behavior, \textit{Correctness} remains insignificant.
\par For translation and optimization where semantic preservation is essential, \textit{Correctness} is the dominant signal: CA@1 and Pass@1, show moderately strong correlations of $r_\rho=0.41$ and $r_\rho=0.59$, respectively. However, since both metrics are binary, tied ranks may cause Spearman correlation to underestimate the impact of \textit{Correctness}. We therefore also report Odds Ratios (ORs)~\cite{repec:tsj:stataj:v:3:y:2003:i:3:p:213-225}. \textbf{A one standard deviation increase in \textit{Correctness} makes a semantics-preserving translation and a semantics-preserving optimization $\mathbf{1.90}$ and $\mathbf{2.62}$ times more likely}, respectively. 
These represent 71\% and 68\% improvements over the strongest baselines, respectively. This gain can be attributed to \sonar's regeneration-based design for \textit{Correctness}: if a summary can regenerate code that is functionally equivalent to the original, it is also more likely to support behavior-preserving implementations in another language or as an optimized variant. 
\par
\begin{figure}[t]
     \centering
  \includegraphics[width=0.45\textwidth]{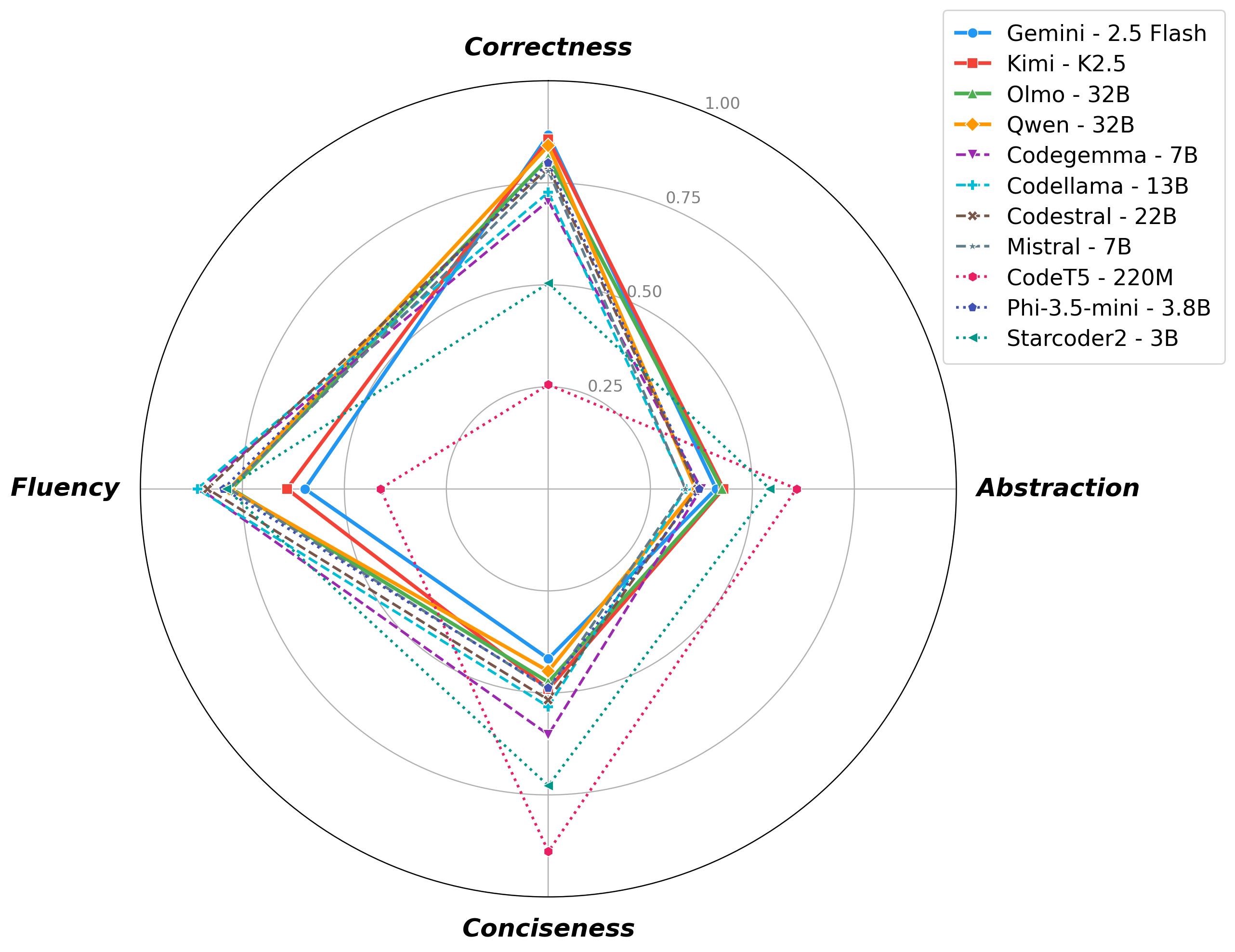}
    \caption{Per-dimension \sonar scores for 11 LLMs.}
    \label{fig:RQ3_Spider_chart}
\end{figure}

 For translation diversity, measured by edit distance, \textit{Abstraction} shows a significant positive correlation ($r_\rho=0.28$), indicating that abstract summaries enhance diversity not only within the same language, but also across languages. Optimization speedup is not strongly predicted by either \sonar or the baselines, though \textit{Abstraction} shows a slight positive trend $(r_\rho = 0.11)$, suggesting fewer implementation details may give the LLM more freedom to find faster alternatives. Finally, \textit{Correctness} is the most significant predictor of success rate in test oracle generation ($r_\rho=0.41$), outperforming the best baseline by 105\%. More importantly, any positive correlation for a reference-based metric is tied to the particular reference used; a different reference could yield a different result. \sonar, by contrast, evaluates the summary under test directly, without any reference, meaning its correlations are genuine signals that the summary itself drives downstream performance. We note that moderate correlations shown in Table~\ref{tab:rq2_table} are expected, as LLM-generated summaries cluster in a narrow range for most quality dimensions (see Fig.~\ref{fig:RQ3_Spider_chart}). 

\par Interestingly, like the baselines, \textit{Fluency} and \textit{Conciseness} show almost no significant impact on downstream tasks, despite being widely valued as important qualities for human developers~\cite{llm-as-ajudge}, thereby revealing a contrast between what human and LLM consumers value in a code summary. This finding is consistent with the capabilities of modern LLMs: their large context windows make conciseness less relevant, while training on large text corpora allows them to better tolerate less fluent summaries. 
\begin{table}[t]
\centering
\scriptsize
\setlength{\tabcolsep}{4.5pt}
\renewcommand{\arraystretch}{1.4}
\caption{Partial Spearman correlation between each SONAR dimension and downstream task performance.}
\label{tab:rq2_partial}
\resizebox{\columnwidth}{!}{%
\begin{tabular}{l|l|l|l|l|l|l}
\toprule
\multirow{2}{*}{\diagbox[width=1.8cm, height=1.0cm]{\textbf{Dimension}}{\textbf{Task}}} 
& \textbf{Retrieval} & \multicolumn{2}{c|}{\textbf{Translation}} & \multicolumn{2}{c|}{\textbf{Optimization}} & \textbf{Test Oracle} \\
\cmidrule(lr){2-2} \cmidrule(lr){3-4} \cmidrule(lr){5-6} \cmidrule(lr){7-7}
& MRR & CA@1 & Edit Dist. & Pass@1 & Speedup & Success Rate \\
\midrule
Correctness & $+0.09$ & $\mathbf{+0.37}^{\boldsymbol{***}}$ & $-0.10^{\boldsymbol{**}}$ & $\mathbf{+0.55}^{\boldsymbol{***}}$ & $-0.05$ & $\mathbf{+0.33}^{\boldsymbol{***}}$ \\
Abstraction & $\mathbf{+0.20}^{\boldsymbol{***}}$ & $-0.07$ & $\mathbf{+0.25}^{\boldsymbol{***}}$ & $+0.08$ & $\mathbf{+0.10}^{\boldsymbol{*}}$ & $-0.15^{\boldsymbol{***}}$ \\
Conciseness & $+0.03$ & $+0.05$ & $-0.01$ & $-0.03$ & $-0.05$ & $+0.06^{\boldsymbol{*}}$ \\
Fluency     & $+0.09^{\boldsymbol{*}}$ & $-0.05$ & $+0.01$ & $+0.06$ & $-0.02$ & $+0.00$ \\
\bottomrule
\end{tabular}%
}
\end{table}
\par \sonar's dimensions are not independent. For example, an overly concise summary may omit behavioral details, hurting \textit{Correctness}, while summaries rich in implementation details may help \textit{Correctness}, but hurt \textit{Abstraction}. To account for these inter-dependencies, we report partial Spearman correlations for each dimension (Table~\ref{tab:rq2_partial}), controlling for the others. With one exception, each significant correlation from Table~\ref{tab:rq2_table} holds in Table~\ref{tab:rq2_partial}, confirming that \sonar dimensions capture task-specific signal beyond their overlap with the others.
\par Although Table~\ref{tab:rq2_partial} reports LLM performance on four tasks and six metrics, the results reveal a broader pattern: \textbf{no single quality dimension is appropriate for all tasks}. 
For behavior-preserving tasks, correctness matters. However, when the goal extends beyond behavior preservation to exploring alternative implementations, abstraction, alongside correctness, helps. But abstraction may hurt tasks that require source-specific details, such as test oracle generation (Table~\ref{tab:rq2_partial}). Again, conciseness and fluency could still be relevant for human consumers in tasks like program comprehension. This is where \sonar's multi-dimensional design becomes valuable: rather than collapsing to a single notion of quality, it enables prioritizing dimensions most relevant to the specific task and consumer, thereby facilitating task-aware evaluation of code summaries.
\begin{rqbox}
\textbf{Findings of RQ2:} At least one \sonar dimension, either \textit{Correctness} or \textit{Abstraction}, significantly correlates with LLM performance for every task, increasing the predictive signal between 1.3 and 14 times relative to the best baseline. What is important in a summary varies with task, and \sonar uniquely captures this variation through its multidimensional design, which existing single-score metrics cannot.

\end{rqbox}
\subsection{RQ3: Evaluation of LLM-based Source Code Summarization using \sonar}
\subsubsection{Experimental Setup} Using the setup described in \S~\ref{sec:Summary_GEn}, Fig.~\ref{fig:RQ3_Spider_chart} illustrates each LLM's average score per \sonar dimension, computed across all summaries generated by that LLM under Vanilla prompting. Mean scores are reported to highlight overall quality trends; finer-grained analysis of score distribution within each model is left for future work.

\subsubsection{Results and Analysis} Fig.~\ref{fig:RQ3_Spider_chart} shows that no single LLM dominates all four dimensions. Most modern LLMs achieve relatively low \textit{Abstraction} scores, typically between 0.25 and 0.50. Smaller models such as CodeT5, achieve higher \textit{Abstraction}, but at the cost of \textit{Correctness}. In contrast, most LLMs score high on \textit{Correctness} (0.75--1.00), with the much smaller 3.8B model, Phi-3.5 Mini, achieving a score of 0.82, comparable to frontier models like Gemini (0.88) and Kimi (0.87). Surprisingly, these SOTA models fall behind open-source mid-sized LLMs in \textit{Fluency}, with Gemini and Kimi scoring 0.14 and 0.10 below the average. To investigate this trend further, we qualitatively analyze the summaries generated by Gemini and find that Gemini often enumerates the implementation step by step in the summary while reusing tokens and keywords from the source code, (Fig.~\ref{fig:motivating_example}). This helps preserve behavior of the code, improving \textit{Correctness}, but it reduces \textit{Abstraction}. Moreover, the abundance of code-related tokens and step-by-step procedural descriptions makes the summaries less natural-language-like, lowering \textit{Fluency}. \textit{Conciseness} varies from 0.42 (Gemini) to 0.90 (CodeT5), indicating substantial differences in average summary length across models. Together, these dimension-specific strengths and weaknesses highlight that summarizer selection should account for the target task and consumer, with preference given to models that perform well on relevant quality dimensions.
\par
\begin{figure}[t]
     \centering
  \includegraphics[width=0.42\textwidth]{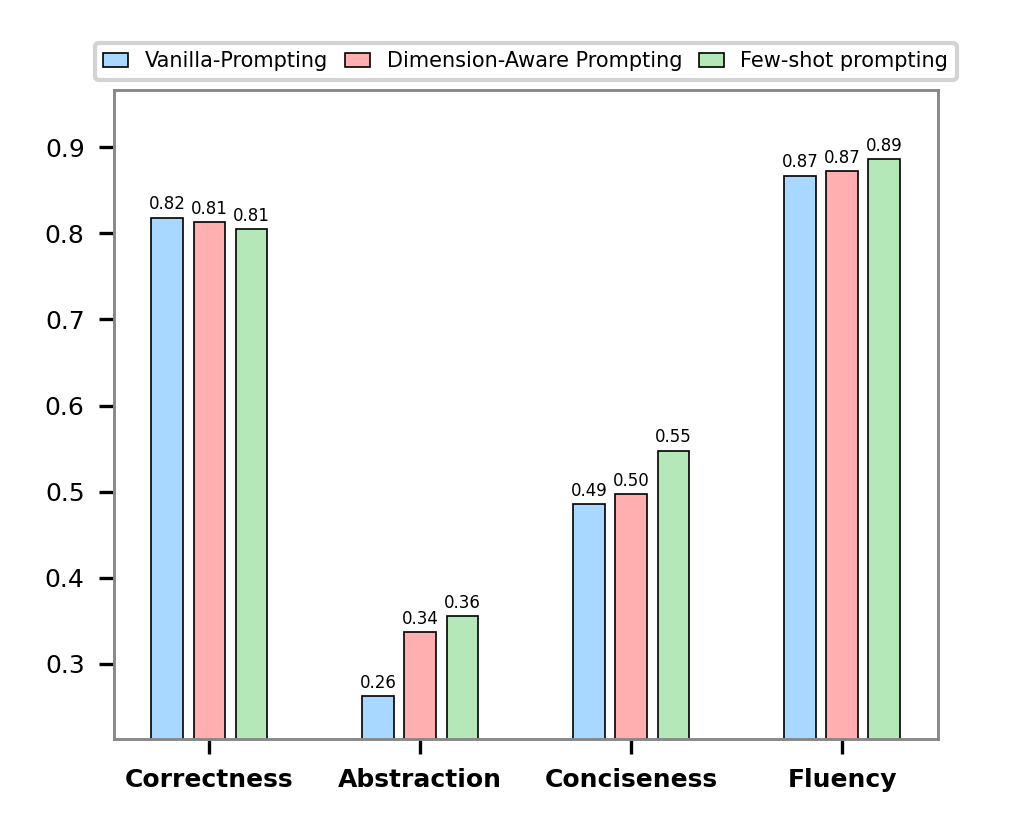}
    \caption{Effect of prompting strategy on \sonar dimensions.}
    \label{fig:RQ3_bar_chart}
\end{figure}

 Motivated by the task-specific role of summary dimensions observed in RQ2 (\S~\ref{sec:rq2}), we analyze how prompting can steer models toward task-aware summaries. Fig.~\ref{fig:RQ3_bar_chart} presents the average score per dimension for all summaries generated under each of the three prompting strategies described in \S~\ref{Section:prompts}. Dimension-aware and few-shot prompting improve nearly all dimensions over vanilla prompting, except \textit{Correctness}, which remains stable. \textbf{\boldmath \textit{Abstraction} improves the most, with gains up to $\approx10$ pp, while \textit{Conciseness} improves by up to $\approx6$ pp.} This demonstrates a lightweight mechanism for task-aware summarization. As an example, for diverse code translation, rather than asking an LLM only to summarize the code, explicitly specifying the relevant dimensions, correctness, and abstraction, or providing a few examples of correct and abstract summaries can steer the summarizer LLM toward summaries that better support the target downstream task.

\begin{rqbox}
\textbf{Findings of RQ3: }LLMs show inherent trade-offs across quality dimensions. Varying prompting strategy provides a practical way to improve task-relevant dimensions, with \textit{Abstraction} showing up to 10 pp improvement, without any additional training. 
\end{rqbox}
\subsection{RQ4: Sensitivity Analysis of \sonar}
\label{sec:rq4}
Since \S~\ref{sec:frameork} defines the four \sonar dimensions through LLM-based interpretation of the summary under test, the scores may be biased by the particular LLM used inside the scorer modules. We test this sensitivity by measuring how much \textit{Correctness} and \textit{Conciseness} scores change under different LLM choices. We exclude \textit{Abstraction} because it already aggregates multiple samples from a multi-LLM pool, reducing sensitivity to any single model. We also exclude \textit{Fluency} since \sonar follows the model choice used in prior work. Thus, we restrict this analysis to the code generator, $G$ used by the Correctness and Conciseness Scorers and the compressor, $P$ used by the Conciseness Scorer.
\subsubsection{Experimental Setup} In our implementation, Gemini-2.5-Flash serves as both $G$ and $P$ in the Correctness and Conciseness Scorers. To analyze sensitivity, we replace Gemini with Qwen2.5-Coder-32B and OLMo-32B-Instruct, two strong open-source LLMs, and recompute \textit{Correctness} and \textit{Conciseness} for 1,000 randomly sampled summaries. We quantify the effect in Table~\ref{tab:RQ4_results} using the Mean Absolute Deviation (MAD) and Spearman correlation relative, $r_{\rho}$, to the original Gemini-based scores, where MAD is defined as:
\[
\text{MAD} = \frac{1}{n} \sum_{i=1}^{n} |z_i^{\text{alt}} - z_i^{\text{Gemini}}|
\]
Here, $z_i^{\text{Gemini}}$ and $z_i^{\text{alt}}$ denote the scores assigned to summary $i$ by Gemini and the alternative model, respectively.
\begin{table}[t]
\centering
\footnotesize
\setlength{\tabcolsep}{5pt}
\renewcommand{\arraystretch}{1.2}
\caption{Sensitivity of \textit{Correctness} and \textit{Conciseness} to the choice of LLM. ``Alt. Model'' denotes the alternative LLM substituted for the original implementation model.}
\label{tab:RQ4_results}
\begin{tabular}{llcccc}
\toprule
\textbf{Dimension} & \textbf{Alt. Model} & \multicolumn{2}{c}{\textbf{Generator ($G$)}} & \multicolumn{2}{c}{\textbf{Compressor ($P$)}} \\
\cmidrule(lr){3-4} \cmidrule(lr){5-6}
& & MAD & $r_\rho$ & MAD & $r_\rho$ \\
\midrule
\multirow{2}{*}{Correctness} 
& Qwen  & 0.042 & 0.891 & \multicolumn{2}{c}{\multirow{2}{*}{N/A}} \\
& OLMo  & 0.055 & 0.878 & & \\
\midrule
\multirow{2}{*}{Conciseness} 
& Qwen  & 0.037 & 0.939 & 0.052 & 0.924 \\
& OLMo  & 0.036 & 0.943 & 0.047 & 0.931 \\
\bottomrule
\end{tabular}
\end{table}
\subsubsection{Results and Analysis}
Table~\ref{tab:RQ4_results} shows that replacing $G$ with a different LLM leads to only small deviations from the original Gemini-based \textit{Correctness} and \textit{Conciseness} scores, with MAD between 0.04 and 0.06. The alternative models also maintain very strong positive correlations with the original scores ($r_\rho = 0.88$--$0.94$) for both \textit{Correctness} and \textit{Conciseness}. This suggests that \sonar is robust to the choice of the code-generator LLM, $G$: the low MAD indicates that scores remain numerically close to the original, while the high correlation shows that the relative ranking of summaries is largely preserved even when Qwen or OLMo is used instead of Gemini. Similarly, replacing the compressor $P$ has limited impact on \textit{Conciseness}, with an absolute deviation of $\approx0.05$ for both Qwen and OLMo and strong correlations of 0.92 and 0.93, respectively. Overall, these results suggest that \sonar is not tied to a single LLM choice; instead, its regeneration-based design remains stable under model substitution, leaving both absolute scores and summary rankings largely unaffected. 
\begin{rqbox}
\textbf{Findings of RQ4: }\sonar remains mostly insensitive to the choice of internal LLM, maintaining very strong cross-model consistency ($r_\rho = 0.88$-$0.94$) with small score deviations (0.04-0.06).
\end{rqbox}
\section{Threats to Validity}
Our findings may be affected by both internal and external threats.
\subsection{Threats to Internal Validity}
The tools and scripts used to implement \sonar and automate our experiments introduce the possibility of implementation bugs. We mitigate this risk by building on widely-adopted tools such as Joern and by rerunning experiments to ensure consistency of results. The validation of \sonar's dimensions relies on human-annotated ground truth, which is inherently subject to inter-annotator variation. To address this, we retain only those labels where both annotators unanimously agree, providing a reasonably sound basis for evaluation. Moreover, to mitigate the risk of author bias influencing the annotation, each annotator independently assigns preferences without access to the other's preferences or \sonar's scores for the pair under evaluation. Although \sonar uses regenerated code to empirically ground each quality dimension, this signal is still an approximation of summary quality. Its accuracy may be affected by the inherent non-determinism of the LLMs used, the influence of their pretrained knowledge, and noise from other framework components, such as the differential fuzzing tool. 
\subsection{Threats to External Validity }The findings on how \sonar dimensions affect LLM performance may be influenced by the choice of LLM used for the task. We mitigate this threat by employing multiple LLMs and reporting the mean correlation across all models. The experiments in  \S~\ref{sec:rq2} rely on MBPP and HumanEval, two established benchmarks for code-related tasks. However, we acknowledge that these benchmarks may not fully reflect the complexity of real-world software, and evaluating \sonar on more complex datasets would further strengthen the generalizability of our findings. Finally, as \sonar currently supports Python only, the applicability of our findings to other programming languages may be limited. The core idea, however, is language-agnostic in principle and can be adapted to support additional languages.
\section{Conclusion}
In this work, we identify abstraction as a new dimension of source code summary quality and present \sonar, a reference-free, multi-dimensional framework that enables task-aware evaluation of source code summaries. \sonar leverages code regenerated from a summary as an empirical signal to assess quality across multiple dimensions, lifting automatic evaluation beyond references, training, and subjective LLM judgment for the first time. Using \sonar, we evaluate code summaries from an LLM's perspective, investigating what aspects of a summary make it useful to an automated consumer. We find that \sonar reliably captures the intended aspects of summary quality and effectively predicts LLM performance across a range of downstream SE tasks. We also find that what makes a summary useful varies with the downstream task and consumer, motivating the need for task-aware summarization. Using \sonar, we evaluate 11 popular LLMs, identify model-specific tradeoffs, and suggest ways to steer code summarization toward specific tasks. As software engineering enters the agentic era, we hope this work broadens the notion of code summarization and its evaluation beyond human needs, opening new opportunities to study summaries from an automated consumer's perspective.\\

\bibliographystyle{IEEEtran}
\bibliography{ref}
\end{document}